# Dark Energy Stars and the Cosmic Microwave Background

G. Chapline

Lawrence Livermore National Laboratory, Livermore, CA 94551

**Abstract**

An initial state for the observable universe consisting of a finite region with a large vacuum energy will break-up due to near horizon quantum critical fluctuations. This will lead to a Friedmann-like early universe consisting of an expanding cloud of dark energy stars and radiation. In this note we point out that this scenario provides a simple explanation for the present day density of dark matter as well as the level of CMB temperature fluctuations. It is also predicted that all dark matter will be clumped on mass scales ~ $10^3$ solar masses.

**1. Introduction**

It was suggested some time ago by Lemaitre and Eddington [1,2] that the observable universe may not have had a singular beginning, but, instead may have originated from a finite size quasi-static condensation of matter that Lemaitre called the "primeval atom". Lemaitre further suggested that this primeval atom was a macroscopic quantum state [3]. More recently it has been suggested that the vacuum of space-time may itself be a macroscopic quantum state [4,5]. Identifying the physical vacuum of space-time as a macroscopic quantum state is very compelling for a number of reasons, but perhaps most notably because it provides a unified explanation for both the observed cosmological vacuum energy [6] and existence of dark matter [7]. The basic idea [8] is that dark matter consists of gravitationally stable quantum droplets – previously referred to as "dark energy stars" - whose interior is like the ordinary space-time vacuum except that the vacuum energy density is much higher. A two-phase model for cosmology arises from a picture of space-time as a mixture of ordinary vacuum and dark energy stars, in somewhat the same way that supersaturated steam consists of a mixture of water vapor and water droplets. It is of course rather natural to imagine that in such a picture the average energy densities of the dark matter and ordinary vacuum might be comparable.

One significant way in which a two-phase model for cosmological space-time differs from supersaturated vapors though is that the long-range gravitational attraction between dark energy stars can lead to the gravitational collapse of dense regions of dark matter. In particular fluctuations in the number density of dark energy stars such that the local dark matter density exceeds the cosmological vacuum energy density can lead to the formation of condensations in the same way that density fluctuations after the era of radiation dominance led to the large scale structures we see today. Moreover one might suppose that when the dark matter dark energy stars begin to overlap almost all of their mass will get converted into vacuum energy. Of course, in the real universe where the energy density of radiation in the later stages of gravitational collapse will be much greater than the mass-energy density of dark matter, a more important question is what happens to this radiation energy. Using general relativity as a guide, the vacuum energy density inside a collapsing cloud of matter must exceed at least ½ of the matter density in order to reverse the gravitational collapse [9]. Therefore a significant fraction of the mass-energy of even ordinary matter must get converted into vacuum energy if during the gravitational collapse sufficiently high densities are reached. The possibility of smoothly converting vacuum energy into ordinary matter was discussed some time ago by Feynman [10]; and by time reversal symmetry the reverse process should also be possible. In the circumstance where the matter undergoing gravitational collapse includes essentially all the matter in the observable universe, we will refer the final state where the matter has been converted into vacuum energy as the "cosmic seed".

We envision that the average vacuum energy density inside this cosmic seed will be on the order of 10 times the energy equivalent of the mass density of closely packed nucleons; i.e. $\approx 10^{16}$ gm/cc. This is also on the order of the maximum possible central density for a neutron star [11], and therefore we adopt this mass density as a reasonable guess as to the minimum density where ordinary matter can easily get converted into the vacuum energy. Indeed during the final stages of the gravitational collapse the mass-energy of stellar matter inside a massive star must get converted into vacuum energy density. In this note we will show that this idea provides a simple explanation for the existence of both dark matter and the CMB. In addition our guess for the vacuum energy density inside the cosmic seed leads us to estimates for the present day average dark matter density and level of CMB temperature fluctuations very close to the values inferred from CMB observations [12].

Our cosmic seed idea differs from the Eddington-Lemaitre primeval atom in that the vacuum energy density inside our cosmic seed greatly exceeds the Einstein critical value. Eddington and Lemaitre assumed that the vacuum energy inside was close to the Einstein critical value in order to make the primeval atom quasi-stationary. Thus in contrast with the Eddington-Lemaitre primeval atom our cosmic seed is not quasi-stable. It would expand exponentially if the vacuum energy inside the cosmic seed were constant; however, conservation of energy prevents this. Instead we propose that the cosmic seed immediately disintegrates into an expanding cloud of dark energy stars with masses $\sim 3M_O$, where $M_O$ is the solar mass. In the next section we argue that this expanding cloud of primordial dark energy stars will evolve into a Friedmann-like universe with radiation and matter consisting of dark energy stars with masses $\sim 10^3 M_O$. After the breakup of the cosmic seed the vacuum energy is unimportant until z ~ 1, when the vacuum energy of the "Λ-sea" begins to be felt. It should be noted that our theory of the early universe has the pleasant feature that there is no "horizon problem" because the expansion did not originate from a singularity. Moreover, what appear as independent parameters in the standard Friedman cosmological model for the early universe [13] – e.g. the ratio of photon density to dark matter density and the level of CMB temperature fluctuations– are actually related in our theory.

## 2. Clumpy nature of dark matter and entropy of the CMB

In contrast with the quasi-stationary Eddington-Lemaitre primeval atom our initial state will be highly unstable because the size of the cosmic seed will greatly exceed the diameter of the de Sitter horizon. This means that the cosmic seed will almost instantly breakup into much smaller quantum droplets due to fracturing of the cosmic seed by quantum critical fluctuations. (In the following we will refer to the primordial dark energy stars initially formed as "quantum droplets" and the dark energy stars formed as a result of condensation of the quantum droplets as "dark matter dark energy stars".) As developed previously [8] in a superfluid model for space-time, the event horizons predicted by general relativity don't occur, but instead are replaced by a quantum critical layers with a finite thickness. Near to these quantum critical layers the superfluid order parameter will exhibit large fluctuations somewhat analogous to the density fluctuations in ordinary gases near to their critical point. For a vacuum energy density equivalent to a mass density of $10^{16}$ gm/cc the nearest quantum crucial layer will less than ~ 4 km away. Thus the cosmic seed for the observable universe will contain an intricate network of quantum critical layers. This network of quantum critical layers will partition the cosmic seed into regions that are by themselves quasi-stable; and therefore it seems reasonable to assume that initially the cosmic seed will break-up into excited quantum droplets with a masses ~ $3M_O$. The problem we face is relating this break-up of the cosmic seed to present day conditions.

The mass $M_*$ of the quantum droplets initially produced in the breakup of the cosmic seed is related to the energy density $\rho_*$ of the cosmic seed by

$$M_* = [2 \times 10^{16} \text{ gm-cm}^{-3} / \rho_*]^{1/2} M_O \quad (1)$$

We are immediately faced with the puzzle though that expansion of a cloud of dark energy stars with mass $M_*$ would lead to a present day density of dark matter that is many orders of magnitude larger than the observed dark matter density. Evidently almost all the mass-energy of the quantum droplets initially produced gets converted into radiation energy. In order to understand how this could be, let us first try to understand how the radiation in a collapsing universe of dark energy stars and radiation might get converted into vacuum energy. A curious property of dark energy stars is that they are very good at absorbing and thermalizing radiation only if the typical quantum energies of the elementary particles are greater than a certain cutoff [14]:

$$h\nu_c \approx 100 (M_\otimes / M)^{1/2} MeV \, , \quad (2)$$

where $M$ is the mass of the dark energy star. This means that if dark matter consists of dark energy stars with a characteristic mass $M_{DM}$ radiation in the collapsing cloud can be efficiently absorbed by these dark energy stars if the ambient radiation temperature exceeds the cutoff for $M = M_{DM}$. For example, if $M_{DM} = 10^3 M_O$ then the radiation in the collapsing cloud will get absorbed when the radiation temperature exceeds a few MeV. The energy storage capacity of a dark energy star with mass $M$ is [8]:

$$U = 50.6 N_* T_{keV}^3 \left(\frac{M}{M_\otimes}\right)^3 M_\otimes c^2 \, , \quad (3)$$

where $T_{keV}$ is the temperature of the dark energy star in units of keV and $N_*$ is the number of different kinds of microscopic degrees of freedom inside the dark energy star. The heat capacity (3) exceeds the heat capacity of the ordinary vacuum by a factor $\approx m_P c^2 / k_B T$, where $m_P$ is the Planck mass. As indicated by Eq. (3), the interior temperatures of primordial dark energy stars will be below ~0.1 keV. However, these equilibrium temperatures may never be reached because of the slowing of relaxation rates for quantum energies below the cutoff $h\nu_c$.

Formally the thermal energy (5) can exceed the zero temperature mass. However, increasing the internal energy density of a dark energy star increases its size, and at some point this would cause the dark energy star to become unstable due to the appearance of internal quantum critical layers, which then cause the dark energy star to break-up into smaller dark energy stars with interior energy densities that are similar to the increased energy density of the larger dark energy star. As time goes on and the ambient temperature exceeds the cutoff frequency (2) for the smaller dark energy stars this process would repeat itself. The process of converting radiation into the mass-energy of dark energy stars will cease when all the radiation has been used up. The cloud of dark energy stars will continue to collapse until they overlap. We now argue that this process can also work in reverse.

Using the cross-section $\sigma = 27\pi(GM/c^2)^2$ for binary collisions between dark energy stars one can show that the collision rate between quantum droplets in the initial matter dominated regime is approximately the same as the expansion time. However, on a similar time scale collections of the initially formed quantum droplets can coalesce to form compact objects via gravitational collapse. The general conditions for this to happen are well known [15,16]. The most widely studied case is when the density fluctuation spectrum has the Harrison-Zeldovich form: [17,18]:

$$\frac{\delta\rho}{\rho} = \varepsilon\left(\frac{ct_0}{R_0}\right)^2 \qquad (4)$$

where $\delta\rho/\rho$ is the fractional density fluctuation within a sphere of radius $R_0$, $\varepsilon$ is the rms metric fluctuation, and $ct_0$ is the horizon radius. If $\varepsilon$ is independent of $R_0$ then Eq.(4) becomes the Harrison-Zeldovich-Peebles spectrum $\delta\rho/\rho \sim k^2$ [17-19]. One can show [16] that, independently of the value of $\varepsilon$, as a result of even a small increase in density within a sphere of radius $R_0$ at the time of the break-up of the cosmic seed $t_0 = (3/8\pi G\rho_*)^{1/2}$, the matter inside this sphere will not expand indefinitely. When the speed of sound within the expanding volume is zero the radius will reach a maximum:

$$R_{\max} \approx R_0\,(\delta\rho/\rho)^{-1}, \qquad (5)$$

and the density contrast at that time will be $\approx 3$. In all cases the volume at the time of maximum expansion will lie inside the horizon. Also in all cases the subsequent collapse time will be less than the expansion time. However only for $\varepsilon \sim 1$ will $R_{\max}$ be close to the Schwarzschild radius. In this case compact objects with a large range of masses will be formed. If $\varepsilon \ll 1$ compact objects will continue to form only as long as the speed of sound remains small; i.e. the expanding cloud remains matter dominated.

When large collections of the initially formed dark energy stars begin to condense there is going to be a large mismatch between their internal energy density and the vacuum energy density inside any much large dark energy star that might be formed as a result of the condensation. In the reverse of what happens in a collapsing universe of dark energy stars and radiation this mismatch is resolved by radiating away most of their mass-energy. As it happens the transition between storing energy as the mass-energy of the initial dark energy stars and radiation can occur very rapidly. If a dark energy star were in thermal equilibrium, its luminosity would be:

$$L \approx 0.4\left(\frac{2GM}{c^2}\right)^2 \frac{(k_B T)^4}{\hbar^3 c^2}\left[\left(\frac{m_P c^2}{k_B T}\right)^{3/2}\left(\frac{h\nu_c}{k_B T}\right)^2 e^{-h\nu_c/k_B T}\right]. \qquad (6)$$

The factor in front of the brackets is on the same order the familiar Stefan-Boltzmann thermal luminosity of a sphere with radius $2GM/c^2$. The first factor inside the brackets represents an enhancement of this luminosity by a factor $\sim 10^{38}$, due to the enormous density of states near to the surface of a dark energy star [8] (This is the dark energy star realization of the "holographic picture" of a black hole [20]). As a result of this enhancement in the luminosity, a condensation of dark energy stars with mass $M$ can radiate away its internal energy on the same time scale, ~10($M/M_*$) μsec, that it takes for the condensation to collapse provided the Wien factor $\exp(-h\nu_c/k_B T)$ is not smaller than about $10^{-23}(M_*/M)^2$. This condition is satisfied for condensation masses $M > 10^8 M_*$. As a consequence, after a time which would allow closely packed condensations of quantum droplets with masses $M > 10^8 M_*$ to form essentially all the mass-energy of the quantum droplets will be converted into radiation, leaving only residual dark energy stars with approximately the same radii as the condensation radii.

As a simple model for the transition between the regime where the dark matter and radiation are decoupled and the high temperature regime where there is strong coupling between the dark matter and radiation we will simply assume that for red shifts greater than a certain red shift, $1+z_r$, the radiation energy is stored as the mass-energy of quantum droplets with mass $M_*$. For $1+z < 1+z_r$, we will assume that the mass-energy of the quantum droplets has been converted into radiation and dark energy stars with mass $M_{DM}$. For the red shift separating these two regimes we will use the value

$$1+z_r = h\nu_c / k_B T_{CMB} , \qquad (7)$$

where $T_{CMB} = 2.73K$ is the present day temperature of the CMB. The radiation energy density as a function of red shift is given by:

$$\rho_{rad}(z) = \rho_* \frac{N(z)}{N(z_r)} \left(\frac{1+z_*}{1+z_r}\right) \left(\frac{1+z}{1+z_*}\right)^4 , \qquad (8)$$

where $1+z_* \approx 10^{13}$ is the red-shift of the initial break-up of the cosmic seed corresponding to the origin of the observable universe. The temperature of the radiation is obtained from the usual relation [13]:

$$\rho_{rad} = N(T) \frac{\pi^2}{30} \left(\frac{(k_B T)^4}{(\hbar c)^3}\right) , \qquad (9)$$

where $N(T)$ is the effective number of elementary particle species present in the radiation. Eq.'s (7-9) yield a present day radiation temperature that is very close to the measured temperature of the CMB if one assumes $\rho_* = 10^{16}$ gm/cc and $M_{DM} = 1500\, M_O$. This model predicts that the CMB originates at a red shift $1+z_r = 1.3\, 10^{10}$. The radiation temperature at this time was 3MeV, so the cosmological production of helium and other light elements should be approximately the same as in the standard model [13].

For red shifts smaller than the red shift $1+z_r$, which marks the beginning of the radiation dominated era of our observable universe, the density of dark matter will be given by

$$\rho_{DM} = \rho^* \left(\frac{M_*}{M_{DM}}\right)^2 \left(\frac{1+z}{1+z^*}\right)^3 . \qquad (10)$$

Again assuming $\rho_* = 10^{16}$ gm/cc and $M_{DM} = 1500 M_O$ we find that the present day density of the primordial dark energy stars is $2 \times 10^{-30}$ gm/cc, which is exactly the present day dark matter density inferred from WMAP [12]. The value of $M_{DM}$ is close to the value that we estimated would survive after most of mass-energy of the initially produced quantum droplets were converted into radiation. Of course a cynic might argue that we have merely chosen $M_{DM}$ to give the observed dark matter density, so this is not a priori prediction. However, the masses of the primordial dark energy stars that are formed by coalescence of the initially formed quantum droplets could in principle be calculated from first principles using a detailed model for the coupled gravitational and radiation dynamics of the quantum droplets in an expanding universe with density fluctuations of the form (4). Thus there is a hope that our model will lead to an a priori theory for the present day ratio of dark matter density to radiation density. It might be noted in this connection that usual way of expressing the specific entropy of the CMB as the number of photons per baryon has no fundamental significance in our theory. Instead what our theory yields is the number of CMB photons per gram of dark matter.

Our simple model does lead to one dramatic independent prediction: dark matter should be clumped on a mass scale $\sim 1000 M_O$. This prediction can perhaps be checked in the near future. For example, when strong gravitational lensing leads to multiple images of the same object, the variations in brightness from image to image is sensitive to the clumpy nature of dark matter [21]. In fact variations in brightness already seen in examples of gravitational lensing with multiple images is suggestive that dark matter may indeed be clumped on some mass scale. Extension of the techniques used in the MACHO project to search for transient micro-lensing by compact objects [22] might also yield direct evidence for our dark matter dark energy stars.

## 3. Why is $\Delta T/T \approx 10^{-5}$ ?

As discussed in the previous section, metric fluctuations in the cosmic seed eventually lead to the production of primordial dark energy stars with masses $\sim 10^3 \, M_O$. The size of these dark energy stars is many orders of magnitude smaller than the overall size of the cosmic seed, which accounts for the overall isotropy and homogeneity of the observable universe. However, density fluctuations during the process of break-up of the cosmic seed will lead to large variations in the densities of the quantum droplets for length scales small compared to the horizon radius. There are two obvious sources of the metric fluctuations in the cosmic seed: 1) density and velocity fluctuations in the collapsing cloud that was the precursor to the cosmic seed would have left an imprint in the cosmic seed, and 2) fluctuations associated with the quantum critical layers. We believe that the second possibility is the more important. Furthermore because the thickness of the quantum critical layers is small compared to the horizon size, the spectrum of metric fluctuations generated by these quantum critical layers should be approximately independent of $k$ for length scales larger than the horizon size $ct_0$. By the well know arguments of Harrison and Zeldovich [17-18] this leads to a density fluctuation spectrum of the form (4). Furthermore, the space-time inside a horizon surface is approximately described by the interior de Sitter metric, where $g_{00}$ decreases from ~1 farthest away from the horizon to nearly zero near to the quantum critical layers. Thus we expect that the spectrum of fluctuations in the density of quantum droplets at the time of break-up of the cosmic has the form (4) with $\varepsilon_0 \sim 1$; i.e. $\delta\rho/\rho \approx (ct_0/R_0)^2$. On the other hand observations of variations in the temperature of the CMB [23,24] as well as theories of the evolution of the large scale structures found in the present day universe [25] are consistent with $\varepsilon \approx 10^{-5}$. Remarkably our theory offers a simple explanation for this difference.

As described in the previous section the CMB originates not at the time of the break-up of the cosmic seed, but at a somewhat later time when the dark matter dark energy stars are being formed. Because the radiation becomes freely streaming when it emerges from the confines of the quantum droplet condensations, the initial density fluctuations will tend to get smoothed out inside the horizon. A reasonable way to approximately evaluate this smoothing is to simply assume that $\varepsilon_0$ is renormalized by the average of the density fluctuations within a light sphere whose radius is the collapse time

$$c\tilde{t} = \frac{2}{3\sqrt{3}}\left(\frac{1+z_*}{1+z_r}\right)^{3/2} ct_0 \ . \qquad (11)$$

Using the Lifshitz result [15] that the density fluctuations during the matter dominated period $z_* > z\, . > z_r$ would have grown by a factor $(1+z_*)/(1+z_r)$ independent of length scale we find that the renormalized value of $\varepsilon$ is given by

$$\varepsilon = \varepsilon_0 \frac{1+z_*}{1+z_r}\left(\frac{4\pi}{3}(c\tilde{t})^3\right)^{-1} \int_{ct_0}^{c\tilde{t}}\left(\frac{ct_0}{R}\right)^2 4\pi R^2 dr = \frac{81}{4}\left(\frac{1+z_r}{1+z_*}\right)^2 . \qquad (12)$$

Using the values for $1+z_*$ and $1+z_r$ implied by eq.'s (7-10) we obtain a renormalized metric fluctuation coefficient $\varepsilon \approx 10^{-5}$. This is an a priori prediction since there are only 2 (and maybe only 1) parameters in our model. Indeed to the author's knowledge this is the first time a cosmological model of any sort has yielded an a priori prediction of this quantity.

The renormalized spectrum of density fluctuations will also be scale invariant in the regime of linear growth. However for large mass scales vorticity may play an important role. If the collapsing cloud that led to the cosmic seed had a non-zero average vorticity, then the effects of this vorticity will be greatly amplified by gravitational collapse. In particular, we expect that an increasing density of vorticity in the collapsing cloud will eventually lead to turbulent velocity flows. In a superfluid model for rotating space-times [26] these turbulent velocities are naturally associated with metric fluctuations because in a superfluid model for space-time the rotation is entirely carried by quantized space-time vortices. As a result for length scales larger than the mean separation between vortices, both the 2-point correlation function for the velocity squared and the metric fluctuations will be nearly scale invariant. The mean-squared velocity and density fluctuation will be related by a universal relation [26]:

$$\frac{\delta\rho}{\rho} = \frac{1}{2}\frac{\langle v^2 \rangle}{c^2} \ . \qquad (13)$$

As noted above, after the formation of the present day dark matter dark energy stars the level of density fluctuations near to a horizon is $\delta\rho/\rho \approx 10^{-5}$. According to eq. (13) these condensations ought to be associated with typical turbulent velocities on the order of 1000 km/sec. Remarkably this is just the typical virial velocity found inside galactic clusters. Thus we are led to the speculation that galactic clusters may contain fossilized evidence of turbulent velocity flows within the primordial superfluid cosmic seed. This may explain, for example, why galactic vorticities are highly correlated within galactic clusters, but only weakly correlated between galactic clusters [27].

## 4. The Flatness Problem

One of the conundrums of Friedmann cosmology is how does the early universe know that the universe is supposed to be flat? In fact in our model the space-time inside both the cosmic seed and the expanding cloud of dark energy stars and radiation is not flat, but has positive curvature. On the other hand, both the dark matter and radiation in our theory evolve smoothly into the conditions seen today.

In our theory the flatness problem is sidestepped in a simple way. Although the spatial curvature inside our cosmic seed is locally positive, the overall curvature of the universe remains flat if we imagine that the cosmic seed was created by gravitational collapse because the average density of dark matter in the overall universe is unchanged. After the time corresponding to redshift $z = z_r$ the densities of both dark matter and radiation in our theory vary with time in a way that is essentially the same as in the standard cosmological model. It is of course an interesting question whether there are any signatures of our supposition that the observable universe arose from a finite cosmic seed. For example, if the cosmic seed rotated, this could show up today as a characteristic large-scale anisotropy of the CMB or a cosmological magnetic field [28].

## 5. Retrospective

Some time ago Zelodvich suggested [29] that the specific entropy of the CMB might be related to the metric fluctuations that for very large scales led to large scale structures such as galactic clusters. Zeldovich imagined that the initial state was a zero entropy state of highly compressed baryons. Thus we share with Zeldovich a conviction that specific entropy of the CMB is intimately related to the level of primordial metric fluctuations and that the energy density of highly compressed baryons is of fundamental significance. However, there are important differences between Zeldovich's suggestions and our theory. First of all, the pressure inside Zeldovich's baryonic fluid is positive whereas the pressure inside our cosmic seed is negative. Also Zeldovich assumes that the CMB arises from the dissipation of sound waves whose wavelength is comparable to the inter-particle separation in his baryonic fluid. In our theory the CMB arises from the conversion of the mass-energy of dark energy stars into radiation. The metric fluctuations associated with this conversion on the scale of the initial separation between primordial dark energy stars are the seed for metric fluctuations at much larger distances, and are naturally scale invariant. Thus in our theory we can not only relate the level of metric fluctuations to the specific entropy of the CMB, but also explain how the metric fluctuations that led to large scale structures were created. Zeldovich, on the other hand, doesn't attempt to explain how the metric fluctuations were created.

The formation in our model of primordial dark energy stars which then act as the cold dark matter responsible for the large scale structure of the observable universe bears some similarity to the proposal by Khlopov, Rubin, and Sakharov [30] that topological defects and "false vacua" in an inflationary universe give rise to massive primordial black holes which then act as the seeds for galaxy formation. Our cosmic seed model also shares some features in common with the "island universe" scenarios [31,32] for the origin of the observable universe. In these scenarios a finite Friedmann-like universe is imagined to originate from a quantum fluctuation in a de Sitter universe with $\rho_\Lambda = \rho_c$ and $\rho_m = 0$. In our model the cosmic seed is imagined to arise from the classical gravitational collapse of dark matter in a Lemaitre universe with $\rho_\Lambda + \rho_m = \rho_c$. However, quantum fluctuations within our cosmic seed do play an essential role in explaining the origin of the CMB and dark matter.

Acknowledgments

This work was performed under the auspices of the U.S. Department of Energy by Lawrence Livermore National Laboratory under Contract DE-AC52-07NA27344. The author is also very grateful for discussions with Pawel Mazur.

References

1. G. Lemaitre, Mon. Not. R. Astron. Soc. **91**, 493 (1931).
2. A. S. Eddington, Nature **127**, 447 (1931).
3. G. Lemaitre, Nature **127**, 706 (1931).
4. G. Chapline in *Foundations of Quantum Mechanics (* World Scientific, 1992).
5. P. Mazur, AIP Conference Proc**. 415**, 299 (1997) (hep-th/9701011).
6. G. Chapline, Mod. Phys. Lett A14, 2169 (1999).
7. G. Chapline, in *Proc Texas Conf. on Relativistic Astrophysics, Stanford, CA 12/12-17/04* (SLAC 2005) (astro-ph/0503200).
8. G. Chapline, E. Hohfield, R. B. Laughlin, and D. Santiago. Phil. Mag. B **81**, pp. 235 (2001).
9. G. Lemaitre, General Relativity and Gravitation **29**, 641 (1997).
10. P. P. Feynman, *Lectures on Gravitation* (Westview Press 2003).
11. M. Nauenberg and G. Chapline, Ap. J. **179**, 277 (1973).
12. E. Komatsu et. al., Ap. J. **S180**, 330 (2009).
13. E. Kolb and M. Turner, *The Early Universe* (Addison-Wesley 1990).
14. G. Chapline, Int. J. Mod. Phys. **18**, 3587 (2003).
15. E. M. Lifschitz, JETP **16**, 587 (1946).
16. G. Chapline, Phys. Rev. **D12**, 2949 (1975).
17. E. R. Harrison, Phys. Rev. **D1**, 2726, (1970).
18. Y. B. Zeldovich, Astr. Astrophys. **192**, 663 (1970).
19. P. J. E. Peebles and J. T. Yu, Ap. J. **162**, 815 (1970).
20. G. Chapline, in *Proc. 12th Marcel Grossman Meeting* (World Scientific 2010) (arXiv: 0907.4397; 1002.4654).
21. P. Schneider, J. Ehlers, and E. E, Falco, *Gravitational Lenses* (Springer 1992).
22. C. Alcock et. al. Ap. J. **499**, 19 (1998).
23. G. Smoot, Ap. J. Lett. **396**, L1 (1992).
24. C. L. Bennett, et. al. , Ap J **436**, 423 (1994).
25. G. Blumenthal, S. Faber, J. R. Primack, and M. J. Rees, Nature **311**, 519 (1984).
26. G. Chapline and P. Mazur (gr-qc: 0407033).
27. M. Schneider and S. Bridle (astr-ph: 0903.387).
28. G. Chapline and P. Mazur, AIP Conference Proc**. 822**, 166 (2007).
29. Y. B. Zeldovich, Mon. Not. R. Astr. Soc. **160**, 1P (1972).
30. M.Yu. Khlopov, S.G. Rubin, and A.S. Sakharov, Grav.& Cosmol. **8**, 57 (2002).
31. L. Dyson, M. Kleban. and L. Susskind, JHEP **0210**, 011 (2002).
32. S. Dutta and T. Vachaspati (astro-ph: 0501396).